\documentclass[review]{elsarticle}

\usepackage{lineno,hyperref}
\usepackage{graphicx}  
\usepackage{dcolumn}   
\usepackage{bm}        
\usepackage{verbatim}   
\usepackage{textcomp}
\usepackage{amssymb}
\usepackage{amsmath}
\usepackage{epsfig}
\usepackage{tabularx}
\usepackage{xcolor}

\journal{NIM A}

\begin{document}

\begin{frontmatter}

\title{Digital Signal Analysis based on Convolutional Neural Networks for Active Target Time Projection Chambers}

\author[dfn]{G. F.~Fortino}
\ead{guilherme.fortino@usp.br}
\author[dfn]{J. C.~Zamora }
\ead{zamora@nscl.msu.edu}
\author[dfn]{L. E.~Tamayose}
\author[ime]{N. S. T.~Hirata}
\author[dfn]{V.~Guimar\~aes}
\address[dfn]{Instituto de F\'isica, Universidade de S\~ao Paulo, S\~ao Paulo 05508-090, Brazil}
\address[ime]{Instituto de Matem\'atica e Estat\'istica, Universidade de S\~ao Paulo, S\~ao Paulo 05508-090, Brazil}

\begin{abstract}
An algorithm for digital signal analysis using convolutional neural networks (CNN) was developed in this work. The main objective of this algorithm is to make  the analysis of experiments with  active target time projection chambers more efficient. The code is divided in three steps:  baseline correction, signal deconvolution and peak detection and integration. The CNNs were able to learn the signal processing models with relative errors of less than 6\%. The analysis based on CNNs provides the same results as the traditional deconvolution algorithms, but considerably more efficient in terms of computing time (about 65 times faster). This opens up  new  possibilities to improve  existing codes and to simplify the analysis of the large amount of data produced in active target experiments.
\end{abstract}


\end{frontmatter}


\section{Introduction} 

In the past few  years,  active-target time projection chambers  (ATTPCs)  have become a quite important and relevant instrument for investigation in  nuclear physics. Several TPC projects have been developed in many laboratories around the world for detailed studies of  nuclear structure, rare decay modes and  particle production with the most exotic nuclei \cite{HEFFNER201450,FURUNO2018215,MAUSS2019498, SHANE2015513, BRADT201765, KOSHCHIY2020163398}. An important advantage of these systems over standard techniques is the  large solid angle coverage and the capability to perform complete kinematic measurements. The active target principle allows the use of a specific gas as a  target medium and   detector  simultaneously with very low-energy detection thresholds. In particular, experiments with low beam intensities can exploit these capabilities for achieving important luminosities due to the relatively large volumes covered by these detectors and the possibility to increase the target thickness without any significant impact in the  angular and energy resolutions. \par
Usually, the detection system used in the TPCs  is based on the  MICROMEGAS (Micro-MEsh GAseous Structures) \cite{GIOMATARIS199629} and GEM (Gas Electron Multiplier) \cite{SAULI1997531} technologies that enable a precise determination of the energy loss from the high gas gain,  and a very good position resolution due to the large granularity of the padplane detector, which generally contains $10^3$--$10^4$ channels. The General Electronics for TPCs (GET) \cite{POLLACCO201881} have been developed to supply the requirements for the  data acquisition with these type of setups with  a large amount of electronic channels. The GET system supports a high-density electronics with front-end data processing that can equip up to 33792 digital channels \cite{POLLACCO201881}. Generally, TPC experiments involve thousands of electronic channels at sampling frequencies in the order of MHz with a maximum data rate of approximately 1.2~Gb/s.  Due to the significant amount of data produced in a TPC experiment, the analysis has to be divided in several stages, where the efficiency and computing time play a critical role. Very efficient algorithms  have been developed in the past few years for performing  pulse shape analysis \cite{GIOVINAZZO201615}, particle tracking \cite{ROGER2011134,AYYAD2018166,LEE2020163840,ZAMORA2021164899},  clustering \cite{DALITZ2019159} and many other important tasks. However, the recent use of machine learning techniques  can bring important advantages for data analysis with TPCs.  \par

Algorithms based on Convolutional Neural Networks (CNN) have been used with a quite success for signal processing and analysis during the last years. For instance, CNNs have shown a great performance for event classification \cite{Acciarri_2017}, pulse shape discrimination, \cite{Holl2019}, pileup correction \cite{FU2018410}, event classification \cite{KUCHERA2019156} and many more \cite{MLSP}. CNNs have the ability to fit multi-variate functions and to learn complex mathematical procedures from a large set of trainable parameters.   Also, the very good time performance of CNN algorithms  over standard methods  provides a clear advantage for the  analysis of a large amount of data. \par

In this work, we present an algorithm for digital pulse analysis using CNNs developed  for experiments with  active targets. We investigated problems in the initial steps of the common pipeline for raw-pulse processing in a TPC experiment, namely   baseline correction, input signal reconstruction, peak detection and peak integration. The CNN architecture employed in our work is explained in Section 2. Section 3 presents a validation of our algorithm using  experimental data. Conclusions and perspectives for new developments with CNN algorithms are discussed in Section 4.

\section{Method}
 Traditional programming generally employs input data and a set of logical rules to obtain a predictive response. Unlike traditional programming, Machine Learning (and also Deep Learning) uses input and response  data (called ``training set'')  to infer the logical rules of a program.  The building blocks of a neural network in Machine Learning are the neurons (or nodes), which are inspired by the brain neural system that contains billions of neurons connected forming a network. These type of algorithms are composed by multiple layers of neurons  that are fully connected to each other. When data are propagated between layers, the response function of a node is given by a multiple linear regression \cite{ML_book}
 \begin{equation}
  y = \sum_i^mw_ix_i + bias,
 \end{equation}
where $x_i$ is the input data, $w_i$ is the weight that determine the importance of a node in a precedent layer and \textit{bias} is an offset which represents how far off the prediction is from the intended value. The connection between adjacent layers is  composed by several of these operations that increase the level of abstraction of an algorithm. Both weights and biases are trainable parameters in the neural network, which are  adjusted to obtain the correct output. \par

CNNs are the most popular Deep Learning networks, which have been designed primarily for the analysis of images. However, in the recent years they have been extensively applied in a range of different fields \cite{cnnreview}. In particular, CNNs have a weight sharing property that reduces the number of trainable parameters and allows to implement large scale networks. This is quite useful for processing a significant amount of data with manifold features, which is indeed the purpose of the present study. \par

\begin{figure}[!ht]
\centering
\includegraphics[width=0.8\textwidth]{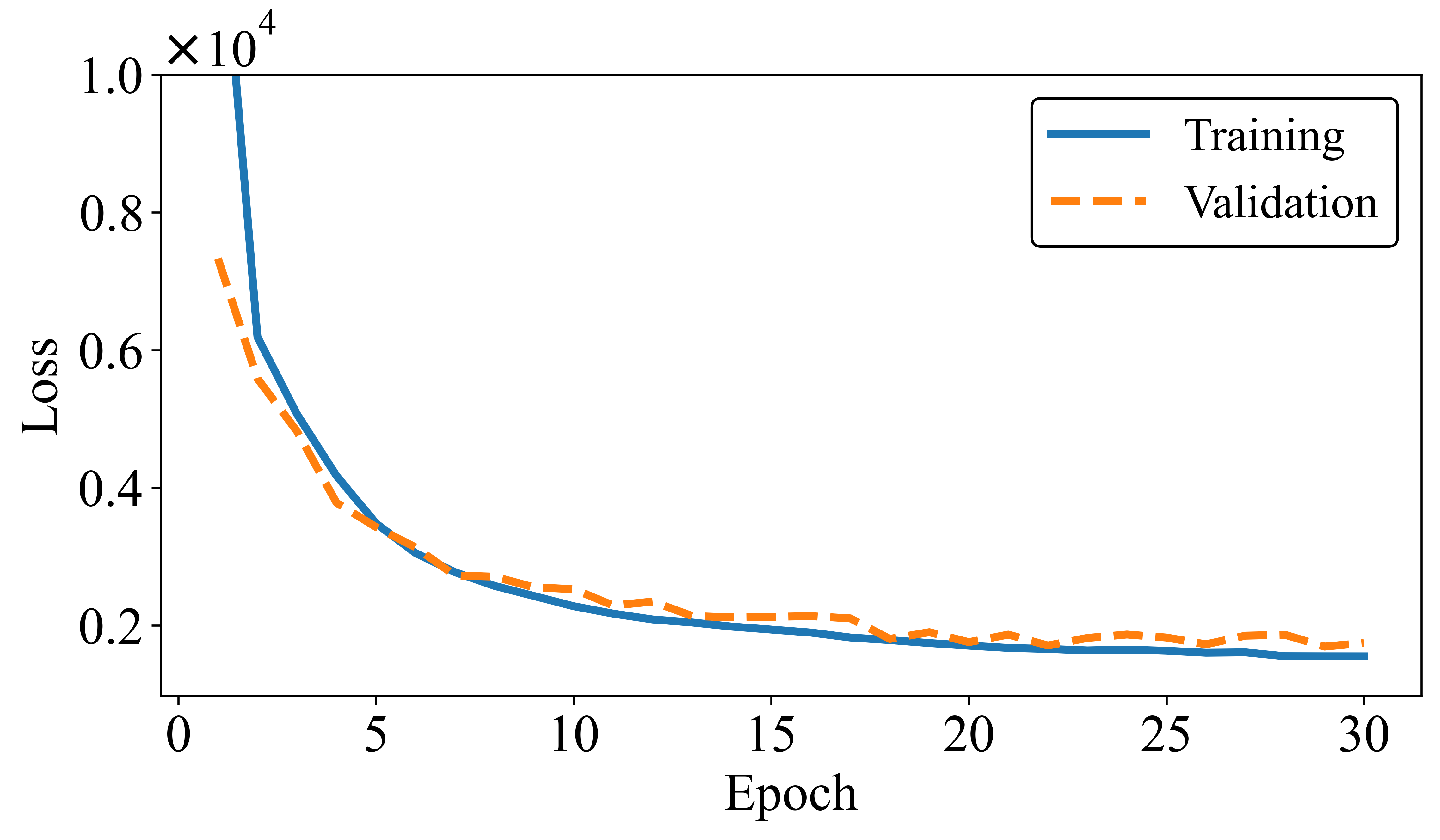}
\caption{\label{loss} (color online). Training and validation Loss curves  of a CNN algorithm of this work.  The model converged approximately after 18 epochs and both training and validation performance remained equivalent. This suggests that the CNN has learned how to solve the problem. }
\end{figure}

In this work, CNN algorithms have been implemented for processing the raw digital signals from a TPC experiment using a GET system. The data employed for this work were measured using the active targets AT-TPC (Active Target Time Projection Chamber) \cite{BRADT201765} and the smaller version (prototype) pAT-TPC \cite{SUZUKI201239}. These devices generate millions of digital signals for each run that contain a complete information about the nuclear reactions in the gas volume. Therefore, the data of only one run can provide sufficient information for setting up the CNN algorithms. The training data set used comprised about $1.6\times 10^5$ pre-analyzed signals randomly selected.  A smaller data set of $4\times 10^4$ pre-analyzed signals was also used as a validation for the trained model. The CNNs were completely developed in Tensor-Flow 2 with more than half a million total trainable parameters.  The  training was GPU (graphics processing unit) based using a NVIDIA Tesla P100, and required about one hour for the whole process. The Loss function was defined as the mean squared error (MSE) \cite{ML_book}
\begin{equation}
 \text{Loss} = \frac{1}{N}\sum_j^N (y_j - \hat{y}_j)^2,
\end{equation}
where $N$ is the number of samples, $y_j$ is the output data and $\hat{y}_j$ is a predicted value. The ADAMAX optimizer \cite{adam} was used  during this procedure with a learning rate of $5\times 10^{-4}$ and batch size of 8. Figure~\ref{loss} shows an example of the training  and validation Loss curves of one of our CNNs.

As can be noticed, in this case the network training reached the best performance at around 18 epochs, where the Loss is well reduced and becomes almost flat. The validation exhibit a similar trend and confirms the success of the CNN training. It is important to mention that the training process needs to be done only once. The trained model can be used  for the analysis of a full experiment, or even for other experiments with the same electronic settings.\par

\begin{figure}[!ht]
\centering
\includegraphics[width=0.8\textwidth]{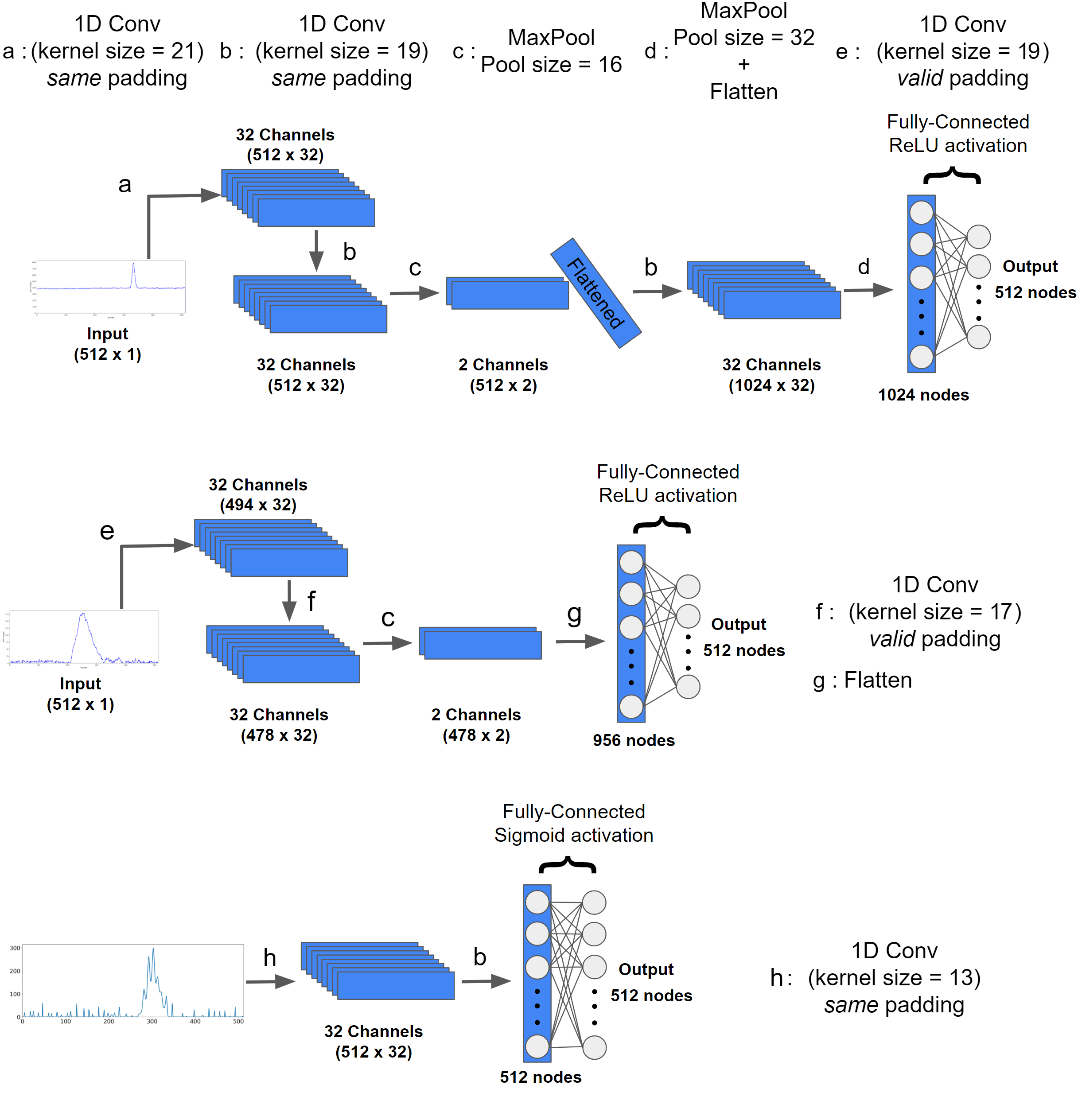}
\caption{\label{structure} (color online). Architecture of the CNN steps of the algorithm developed in this work. The diagram on the top corresponds to the CNN that predict the baseline of a raw signal. The next step is a CNN that performs a signal deconvolution (middle). Finally, the peak centroids and integrals are extracted by a CNN that performs a segmentation of the signal (bottom). The three CNN steps were built to be connected and make the full analysis of a raw signal.   }
\end{figure}

The algorithm developed in this work is divided in three main steps: baseline correction, deconvolution, peak detection and integration. The architecture of the CNNs is presented in Figure~\ref{structure}. The CNNs have a similar structure to the ones used for image classification.  However, in the present case the input data is a one dimensional array of size $512$ that corresponds to the number of time buckets of a digital signal. The output of the baseline and deconvolution CNNs is a signal of the same size as the input data (see Figure~\ref{structure}). Nevertheless, the output of the third CNN is a classification score map, with values between 0 and 1 at each time bucket (the closer to 1 the more likely the bucket corresponds to a peak).

Typical CNNs consist of a number of convolutional layers, eventually intercalated with Max Pooling layers, and followed at the end by a number of fully connected layers. A convolutional layer consists of $k$ filter masks and it computes the convolution operation between the input array and each filter mask (also called kernel). The output of a convolutional layer is, therefore, a set of $k$ filtered signals. The resulting filtering effect depends on the mask size and values; a filter may enhance some features of the input signal. Thus, the output of convolutional layers is commonly known as feature maps. For instance, in the baseline correction network, the input is of dimension $512\times 1$ meaning one signal with 512 buckets, while the output of the first convolutional layer is of dimension $512\times 32$, meaning 32 signals with 512 buckets (or a signal of size 512 with 32 channels). The number of  filters, $k$,  is an empirical value that is chosen to improve the performance of the CNN.\\
The size of the kernels were chosen between 17 and 21 buckets to fully cover a peak region. Note that due to the finite size of the signals, the convolution operation can not be applied to the points in its extremities (first and last time buckets). In convolutional layers, if the output signal size must be equal to the input signal size, then padding (artificially adding sufficient number of points in both extremities, usually filled with zeros) is employed. Without padding, we have outputs of reduced size as the one in the second architecture of Figure~\ref{structure} (since the kernel size is 19, one can not apply the convolution operation on 9 points in each extremity and, thus, the resulting signal is of size $494=512-18$).\\
Max Pooling is another common operation in CNNs. Its main effect is to reduce the dimensionality of data. It consists of a sampling process based on a sampling window with a certain pooling size, keeping the maximum value (hence, the name Max Pooling) in each pool. Pooling can be applied with respect to the time dimension of signals, as well as with respect to the channel dimension. In our case, we apply, for instance , a Max Pooling of size 16 to reduce the number of channels from 32 to 2. This reduces the number of parameters, and, as a consequence, it shortens the training time. At the end of the network, a couple of fully connected layers are included. Between the last convolutional layer and the first fully connected layer a fattening operation is applied to convert a multidimensional feature map into a unidimensional array. We used ReLU (rectified linear activation unit) function [$g(x) = \max(0, x)$] to guarantee a minimum value at zero and to prevent gradient saturation.  The ReLU function is also a non-linear activation that helps the network learn a non-linear operation, as the ones required in this study. Finally, in the peak detection and integration CNN, the nodes in the fully connected layer are followed by a sigmoid activation function which confines the node output to the interval $[0,1]$. Accordingly, in this case we replaced the MSE loss function by a binary cross-entropy  funtion \cite{ML_book} . \par

The three CNN steps explained above are designed to be connected in a sequential way to fully analyze the raw signals and to reconstruct point clouds in a much faster way than the traditional deconvolution methods used so far. A few examples of pulse processing for each CNN are presented in the next section.

\section{Validation with Experimental Data}

\subsection{Baseline correction}
The baselines of the GET digital signals in a  TPC experiment are quite complex since their shape can vary for every channel and also on event-by-event basis. On the one hand, coherent fluctuations  can be produced by charge induction on a certain section of the padplane by capacity coupling through the mesh. On the other hand, the circular buffer memory of the GET electronics can generate systematic effects in the baseline shape that are repeated periodically \cite{GIOVINAZZO201615}. In order to correct for such effects, usually extra FPN (fixed pattern noise) channels are used to estimate an average coherent noise of the electronics on event-by-event basis \cite{POLLACCO201881}. However, the storage of extra FPN channels may represent the same amount of data as the signal channels. Another option is to estimate the baseline shape directly from the raw signal. This can be done, for instance, with a moving average or with a Fourier transform \cite{joshPhD}. With similar good results than the latter method, the peak clipping algorithm is a very successful way to estimate the baseline of the raw signals \cite{MORHAC1997113}. In this work, the training data set for the CNN algorithm was obtained with the peak clipping algorithm implemented in the ROOT libraries \cite{citeulike:363715}. \par

\begin{figure*}[!ht]
\centering
\includegraphics[width=0.49\textwidth]{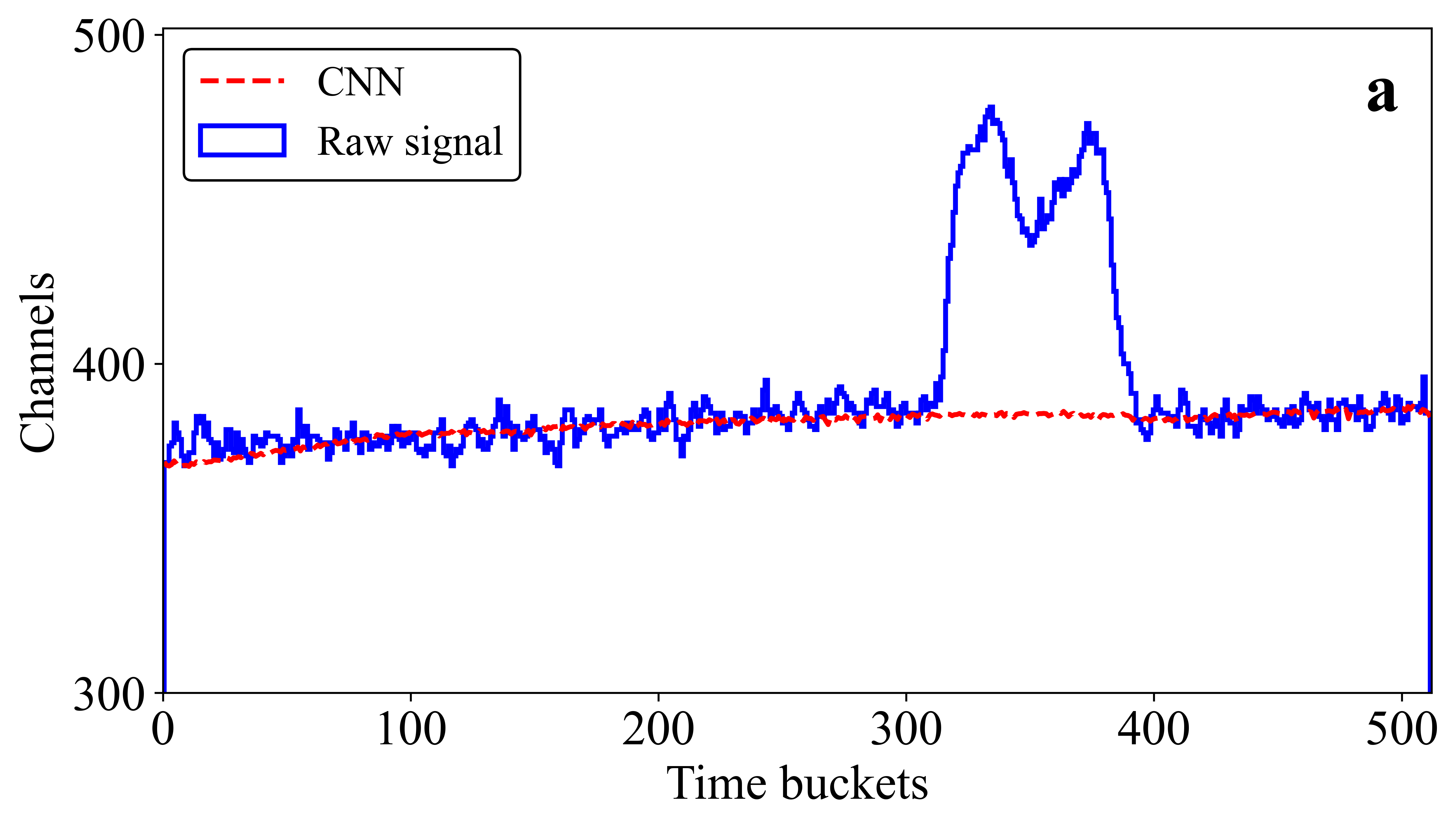}
\includegraphics[width=0.49\textwidth]{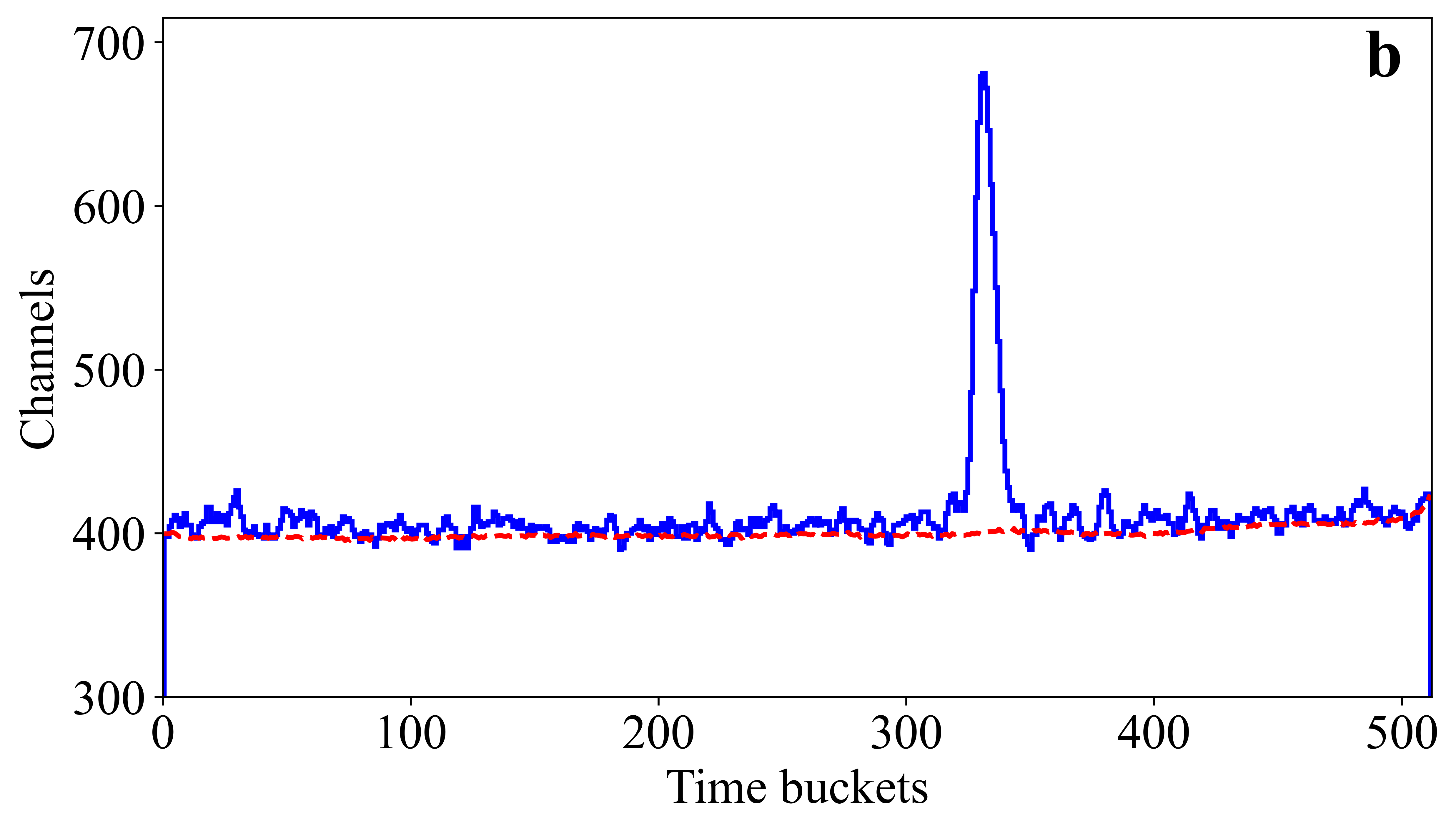}\\
\includegraphics[width=0.49\textwidth]{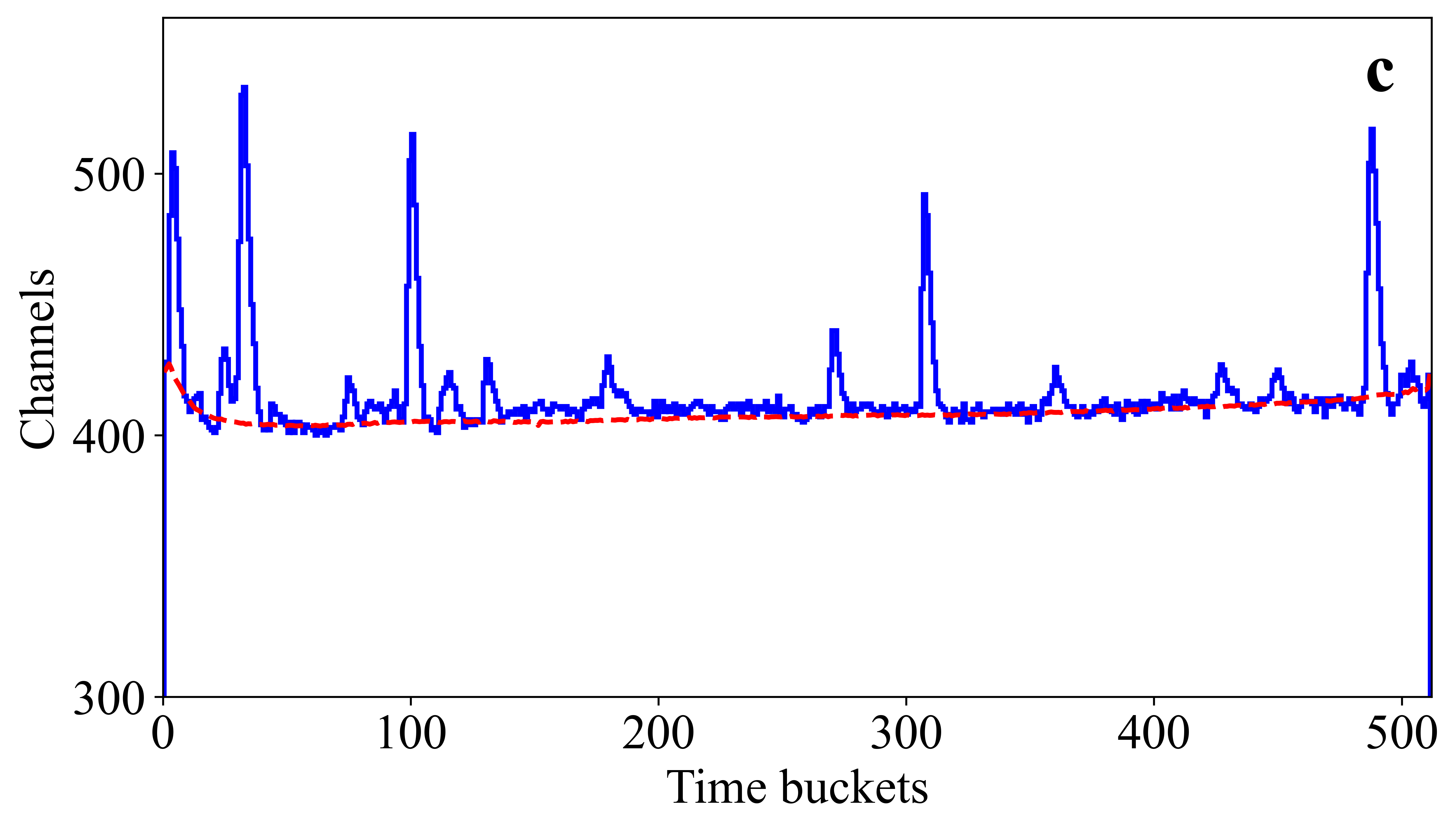}
\includegraphics[width=0.49\textwidth]{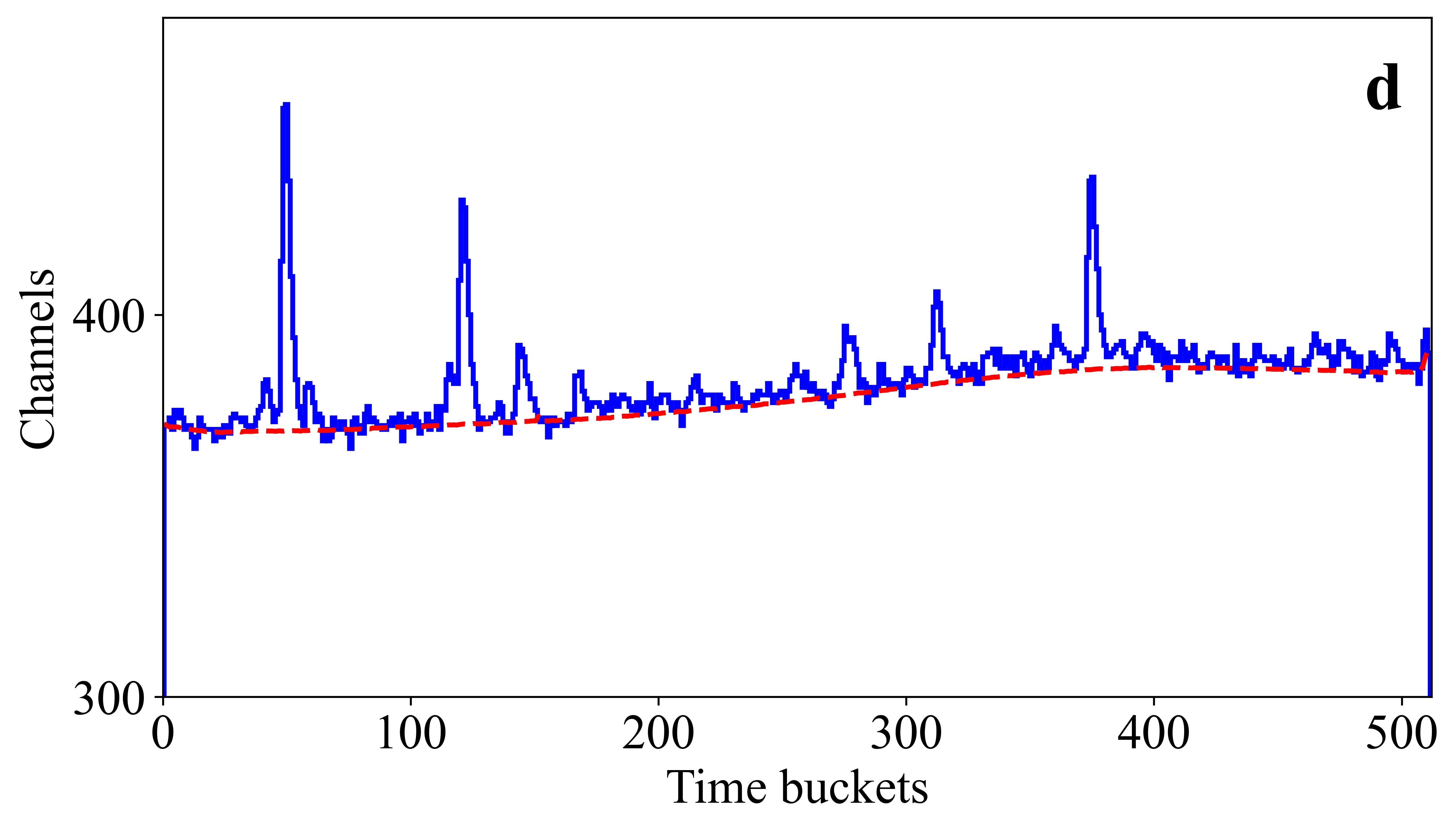}
\caption{\label{baseline} (color online). Examples of baseline prediction using the CNN algorithm (step 1) developed in this work. The output is a smooth distribution that reproduces successfully the oscillations of the experimental baseline signals. The algorithm allows the identification of peaks with amplitudes up to 10 channels.  }
\end{figure*}

The baselines (divided in 512 bins) used as a  training data set in the CNN were smooth   distributions that average the amplitudes without including the peak regions. In other words, the network was trained to find values far off a coherent behavior and  reject them from the smoothing. Figure~\ref{baseline} shows a few examples of the predicted baseline  for different cases. 
As can be seen, the baseline reproduced by the CNN algorithm is remarkably good. The non-linear low frequency oscillations of the baseline is very well reproduced, including  sudden jumps at the spectrum edges. The peak rejection is clearly  one of the strongest feature of the network, that is able to discriminate peaks above the baseline with amplitudes up to 10 channels, which is slightly above of the noise level (approx. 5 channels). The relative error of the CNN baseline predictions with respect to the output of the peak clipping algorithm  is smaller than  1\%.

\subsection{Hit signal reconstruction}
The signals of a TPC detector correspond to the energy loss of the charged particles in the active volume. Each particle produced by nuclear reactions interacts with the gas volume producing ionization electrons. These electrons  drift towards the padplane due to a constant electric field applied in the TPC. The high sampling frequency of the GET system allows the reconstruction of the electrons drift time from the digital pad signals. However, depending on the incident direction of the particle, with respect to the electric field, the charge projection on the individual pads can have different distributions \cite{GIOVINAZZO201615}. For example, if the particle direction is perpendicular to the electric field, the signal is a well defined peak  because the electrons arriving in this pad were produced in a narrow region of the drift axis.  But if the particle direction is parallel (or antiparallel) to the electric field, the projected charge on a pad  is a wide signal in the sampled time. In this case, the electrons were produced at different positions along the drift axis, and generate a signal pileup of the particle hits.  The  pileup effect has been studied in other works by using a signal deconvolution with a Fourier transform and assuming a certain electronics response function \cite{GIOVINAZZO201615}. In this method, the projected charge signal is decomposed into different contributions allowing the identification of the time and charge of each particle hit. The Gold deconvolution algorithm \cite{gold1964iterative} is also a very powerful method to decompose the raw signals in a superposition of Gaussian functions \cite{MORHAC20111629}. The advantage of this algorithm is that the solution is always positive without problems of numerical fluctuations.\par

The CNN was trained with the analyzed raw signals using the Gold deconvolution algorithm implemented in the ROOT libraries. The Gaussian parameters and number of iterations were adjusted to reproduce the resolution of the peaks without pileup. This procedure gave us a confidence to reconstruct the hit distribution on each pad sensor. \par

\begin{figure*}[!ht]
\centering
\includegraphics[width=0.49\textwidth]{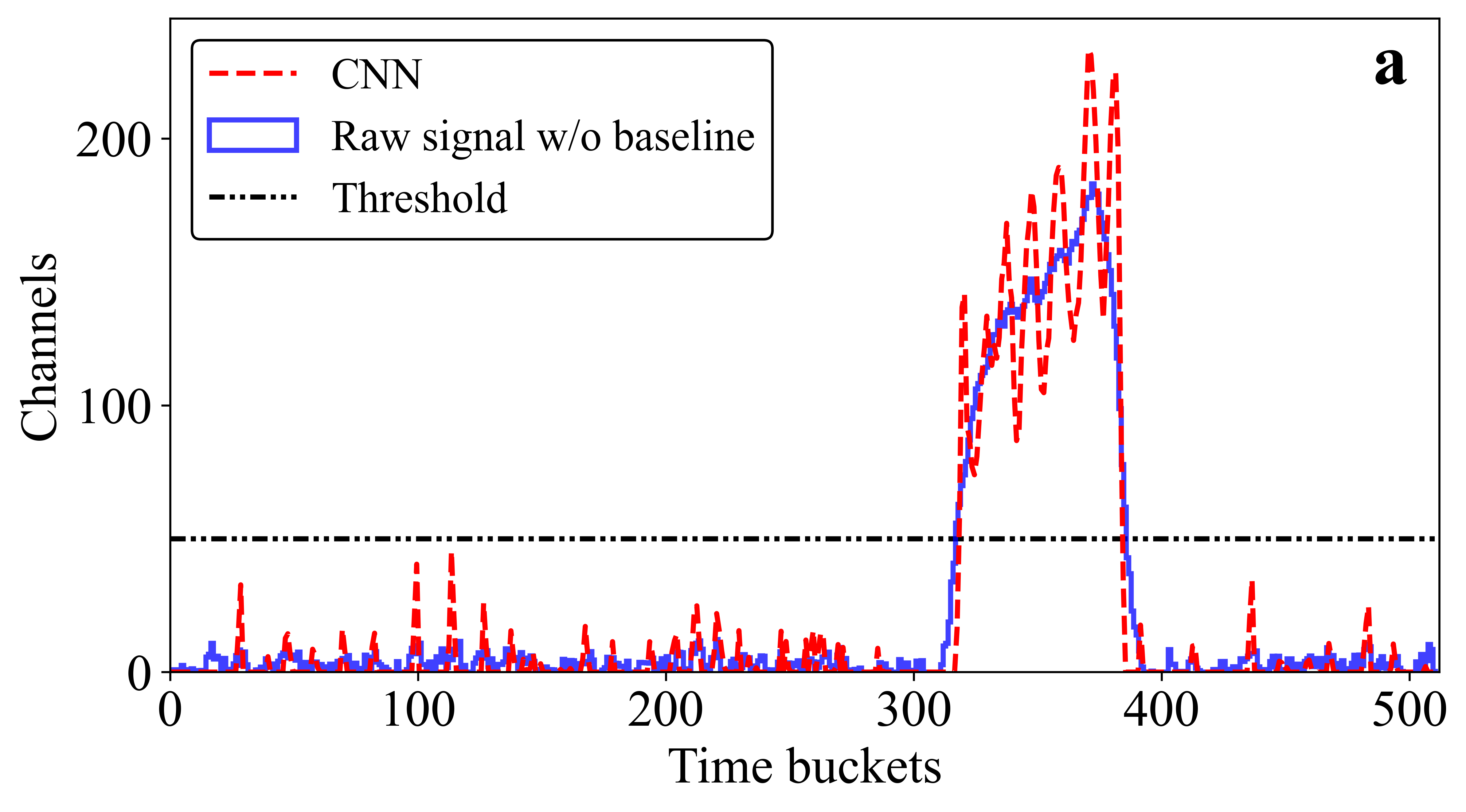}
\includegraphics[width=0.49\textwidth]{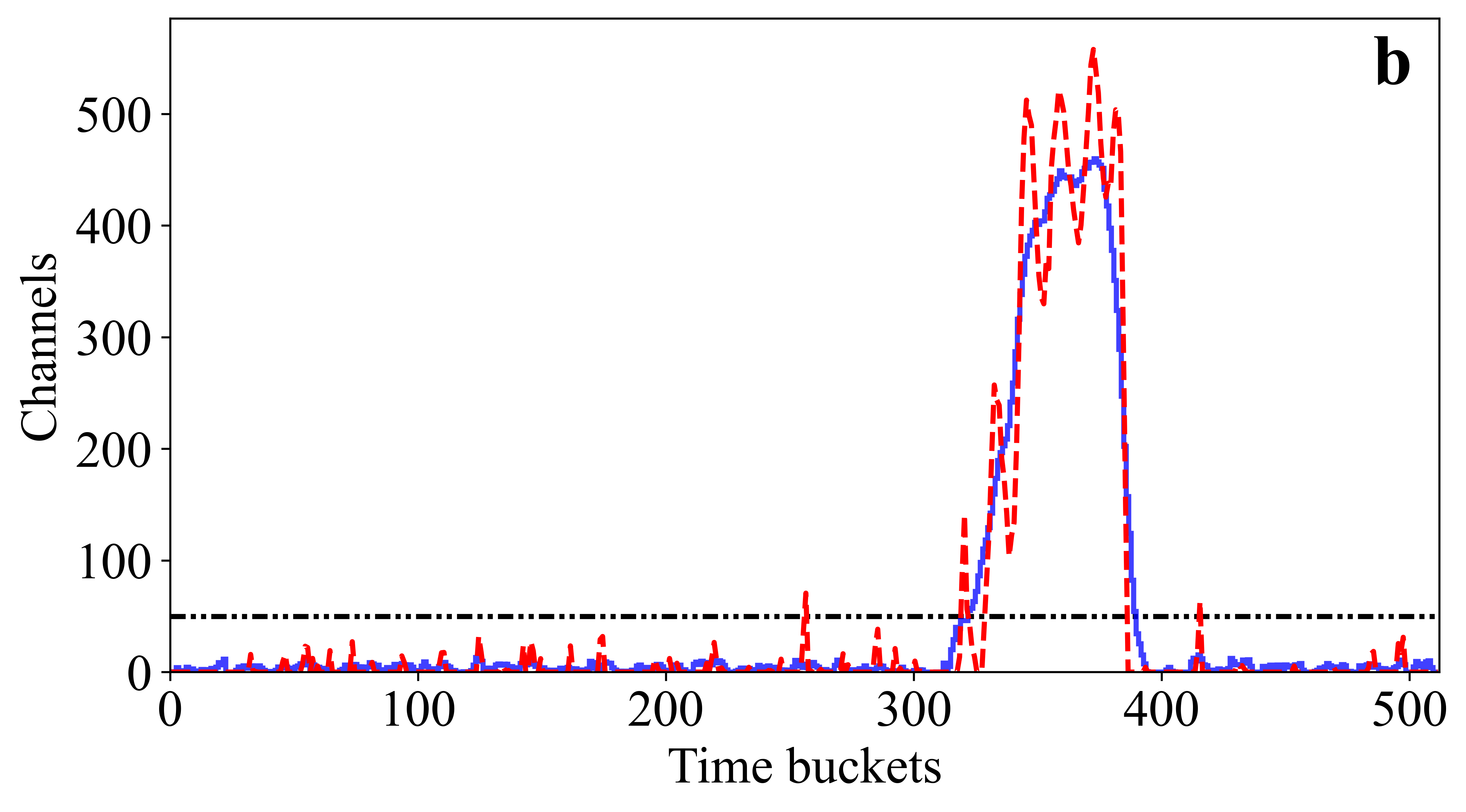}\\
\includegraphics[width=0.49\textwidth]{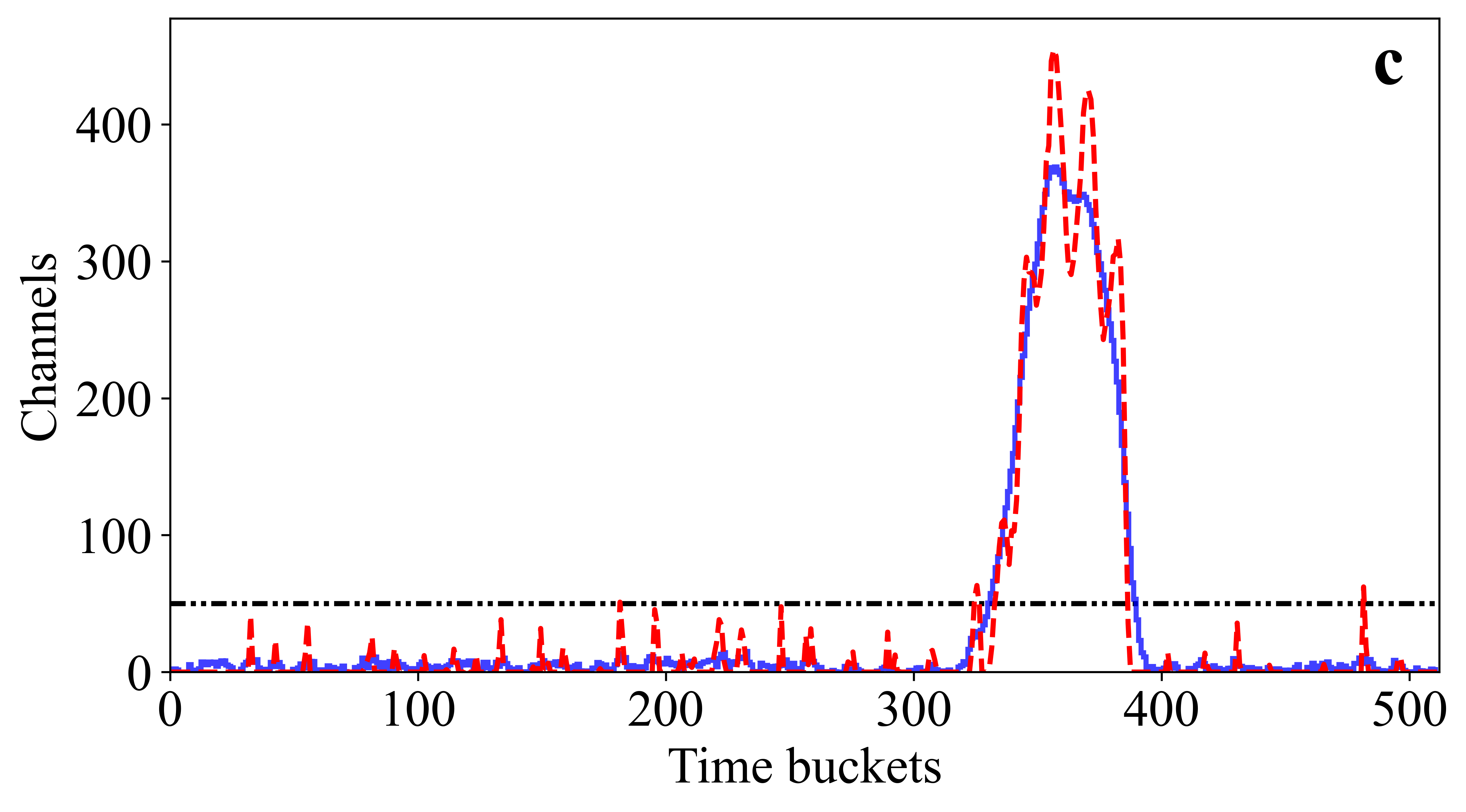}
\includegraphics[width=0.49\textwidth]{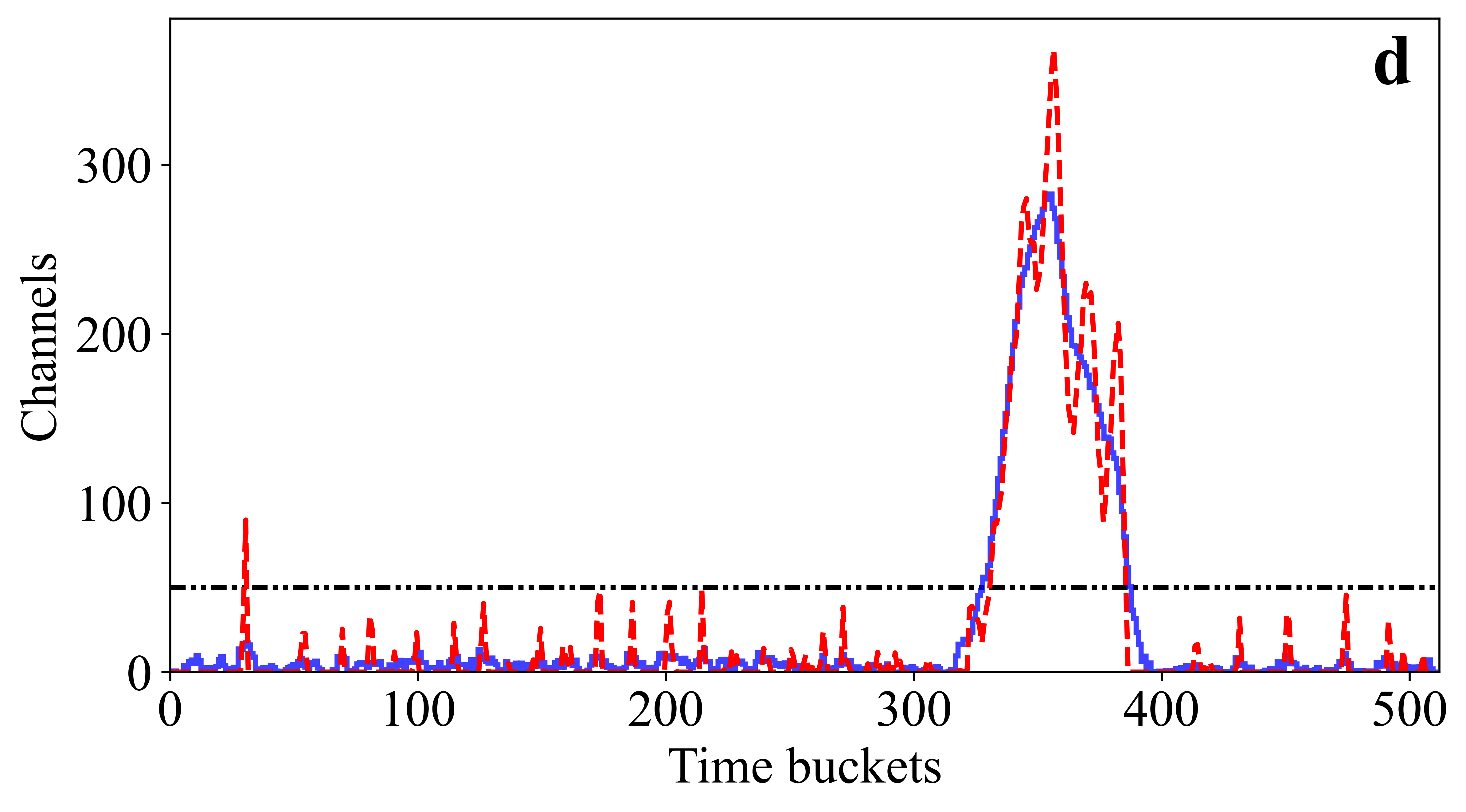}
\caption{\label{deco} (color online). Examples of the CNN prediction for the peak deconvolution (step 2) developed in this work. The input signals without baseline are decomposed in several Gaussian peaks with a certain width. This procedure is important for the analysis of raw signals with an extended pulse shape (pile up). The horizontal dashed-dotted line is the threshold used in each case to accept the peaks. }
\end{figure*}

Figure~\ref{deco} shows a few examples of the CNN prediction with the respective raw signals. As can be seen, the raw signals were deconvoluted in a superposition of Gaussian peaks that conserve the integral of the original spectrum. The CNN algorithm have learned how to perform a quite complex signal processing and still preserving important properties such as the positive solution and integral conservation. It is also interesting to note that the CNN is able to deconvolute the very small peaks and residuals after the baseline subtraction. As most of these peaks are $\delta$ electrons and noise, a sharp threshold is used to accept only the most intense peaks originated in a nuclear reaction. This type of signal processing is very useful for the data analysis in TPCs with a low granularity such as the pAT-TPC \cite{SUZUKI201239}, where the pileup in the charge projection is most common. The uncertainty  associated with the CNN prediction relative to the output from the Gold deconvolution algorithm is below  6\%. However, the output of this CNN step consists just of a transformation of the pulse shape. Thus,  no information about the different Gaussian components is extracted because the network was trained to provide only a ($512\times 1$) output array. In the future, we have plans to extend the code to extract multiple output arrays that contain the different components of a raw signal.

\subsection{Peak centroid and integral}
The two CNN steps, described in the previous sections,  were able to subtract the baseline and to deconvolute the signals. However,  it is still necessary to obtain the time and charge of each peak in order to reconstruct a point cloud. Therefore, the third CNN step corresponds to the peak detection and integral.\par
The main problem is to find a peak centroid value into  512 possible positions. This corresponds to a well-known problem of imbalanced classification in Machine Learning, where only one channel have the largest  probability  to be selected over the rest of the sample set. In order to solve this problem, an image segmentation was applied to the raw signals. This procedure is commonly used in computer vision algorithms \cite{aly2011research}.   The idea is to divide the input signals in segments (or windows) map to train the network. Now, instead having only one point, we employ a window (with their respective values) around the peak centroid, which enrich the features map to the CNN. The training set was obtained with the SciPy \cite{scipy} library that determines the peak centroids of the deconvoluted signals with a good precision. Each peak centroid was associated to a spectrum segment that contained the most probable position  to find a peak. The windows were adjusted to be consistent with the peak width without pileup. An advantage of this training method is the possibility to obtain the amplitude values around a peak position, which immediately provides the information to extract  integral of the peak. The CNN response is an output map with values between 0 and 1 for the 512 time buckets. However, only the candidate peaks with output larger than 0.5 are considered as a full valid window. \par

\begin{figure*}[!ht]
\centering
\includegraphics[width=0.49\textwidth]{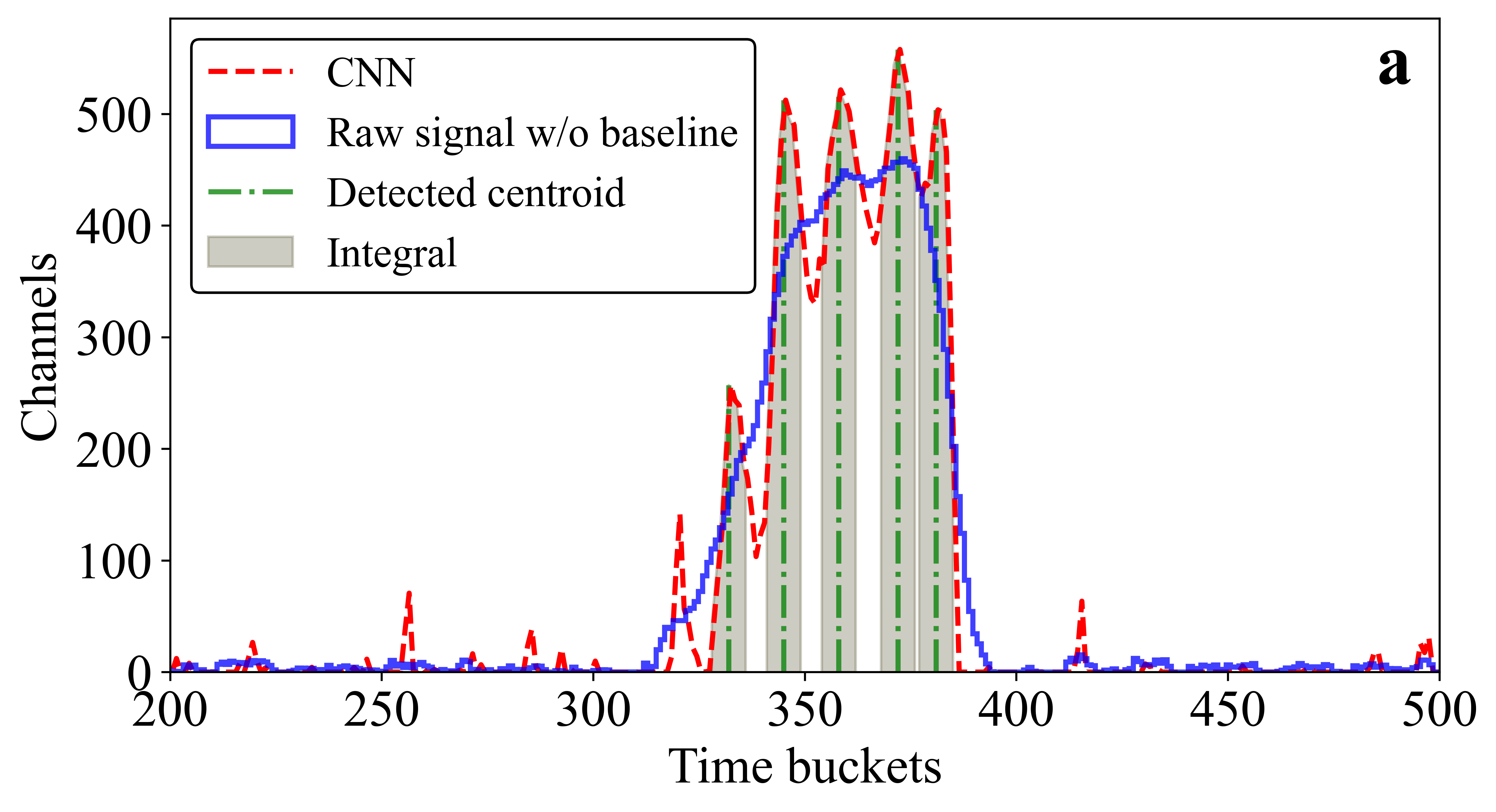}
\includegraphics[width=0.49\textwidth]{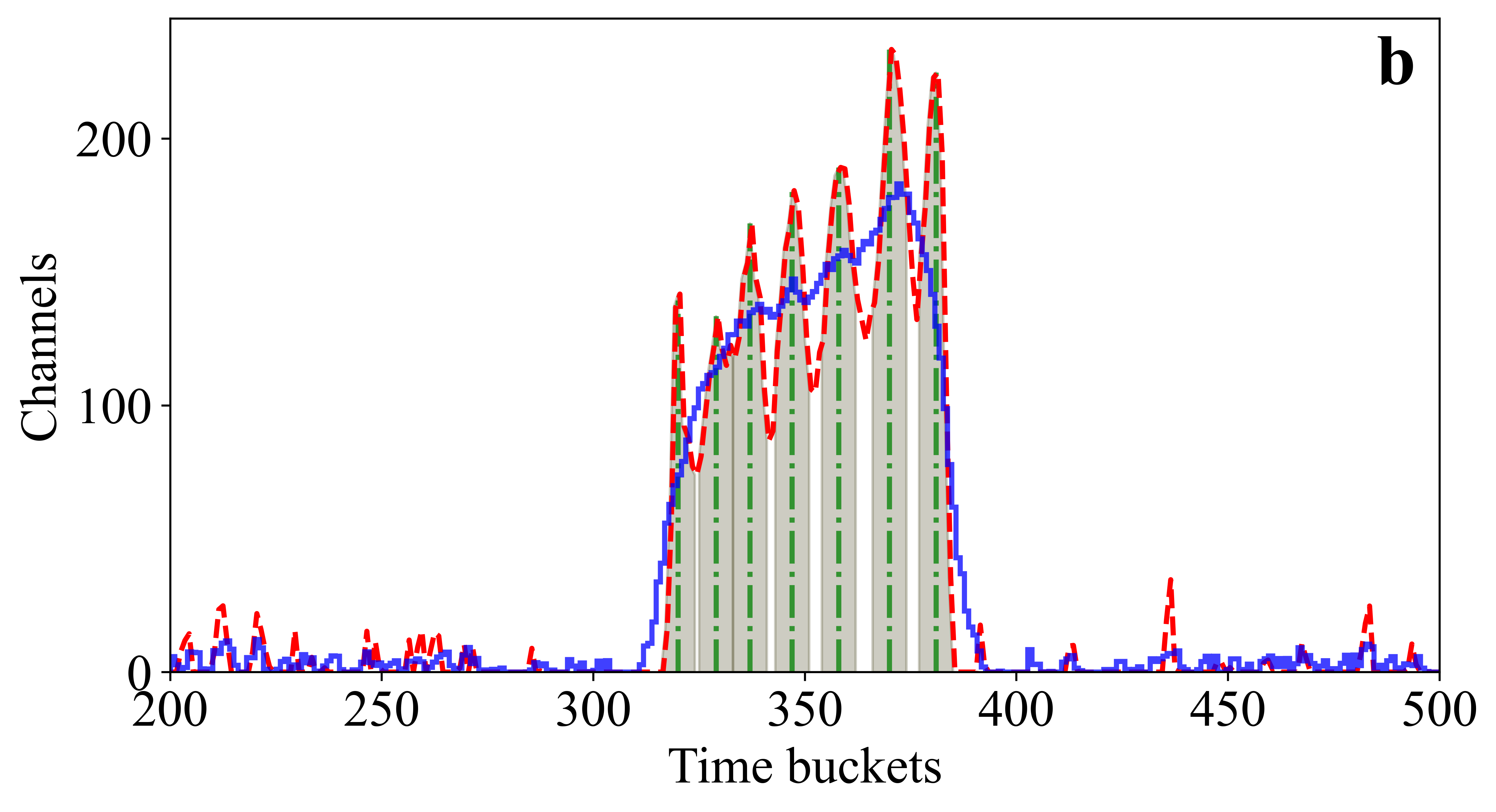}\\
\includegraphics[width=0.49\textwidth]{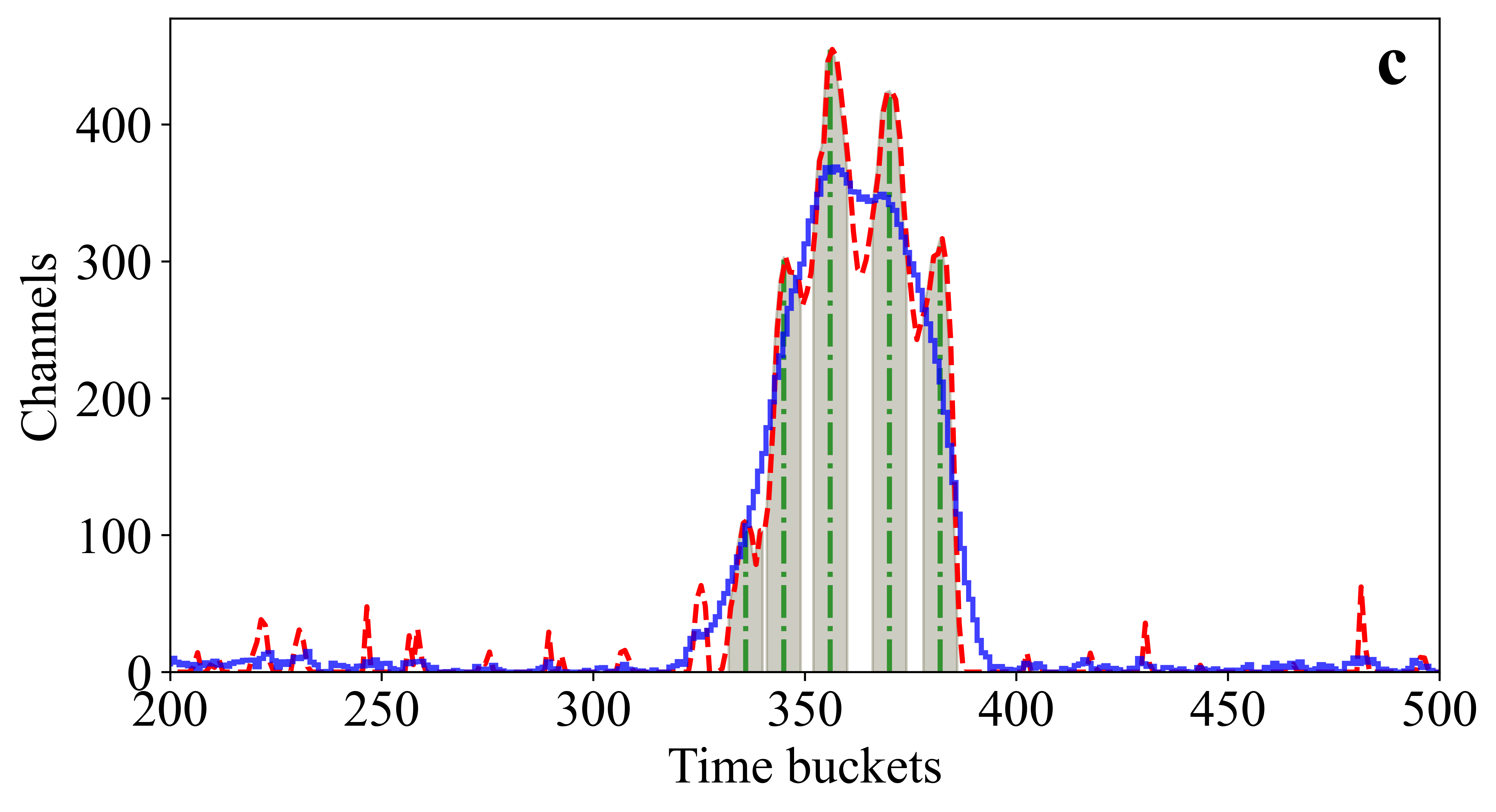}
\includegraphics[width=0.49\textwidth]{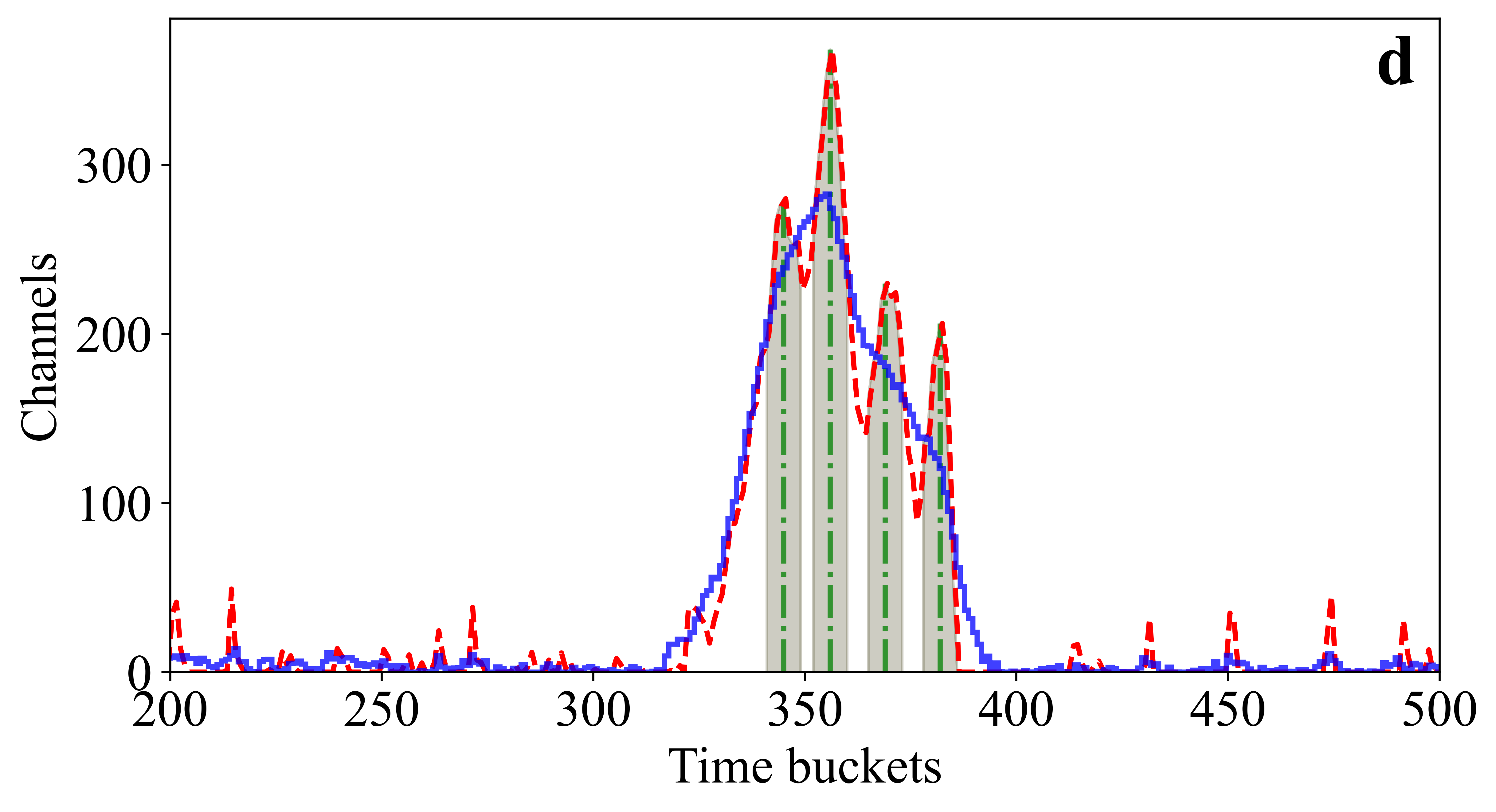}
\caption{\label{peaks} (color online). Example of the CNN prediction for the peak detection and integration (step 3) of our algorithm. The CNN input are the deconvoluted signals. The response function is a map of  signal segments that can contain a peak. Centroids and integrals are directly obtained from the information of these segments.  }
\end{figure*}

Figure~\ref{peaks} shows a few examples of peak detection and window integral for deconvoluted signals. The peak detection and integral for a single peak spectrum is rather simple with the present method. Therefore, we focus mostly in the analysis on the wide signals with pileup. As can be observed, the peak detection using the deconvoluted signal is very precise and the integral covers most of the area of the raw signal spectrum. The accuracy of the CNN to detect a peak in a window is about 93\%. The precision  of the peak position is better than two time buckets. However, the algorithm suffers a drawback for the peak detection of areas without visible peak maximum. For example, the algorithm was not able to find a peak at the  lower region of Figure~\ref{peaks}(d), where a peak maximum is not clearly visible. As it was previously commented, a possible solution for this problem would be to decompose  a signal image in several components.

\subsection{Point Clouds reconstruction}
\begin{figure}[!ht]
\centering
\includegraphics[width=0.8\textwidth]{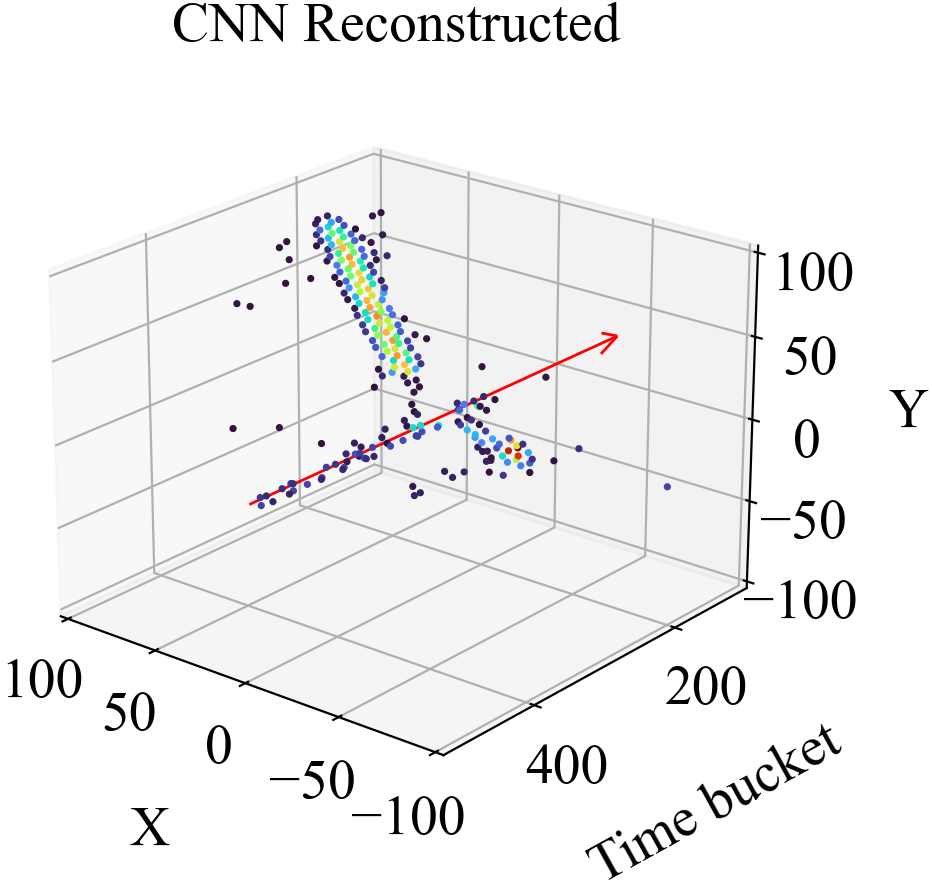}
\caption{\label{pointcloud} (color online). A point cloud fully reconstructed with the CNN algorithm. The red arrow is the beam direction. The other tracks with the same vertex are scattered particles of a reaction. The $(X,Y)$ coordinate corresponds to the position on the padplane, while the time axis is extracted from the peak centroids. The color scale corresponds to the integrated charge of each hit. }
\end{figure}

The architecture of the present CNNs allows to connect each step in a sequential way to directly extract the positions, drift times and integrated charges from each raw signal. The analysis is performed in data subsets to preserve the event structure.  With this information is possible to reconstruct the point clouds of the nuclear reaction events.   Figure~\ref{pointcloud} shows an example of the CNN reconstructed point cloud. As can be seen, the beam track (red arrow) is parallel to the drift time coordinate. The points along the beam track were successfully extracted from the deconvolution step. The color of the points represent the integrated charge of each signal.\par

The coupled algorithm is able to fully analyze $2\times 10^5$ signals and to reconstruct the point clouds in about 23 seconds. However, the same analysis expended 25 minutes using the peak clipping and Gold convolution algorithms.   This result shows the great capability of the CNN codes that can produce the same outputs in a considerably shorter time.

\section{Conclusions}
A CNN algorithm for digital signal analysis of active target experiments was developed in this work. The code was divided in three steps: baseline correction, signal deconvolution and peak detection and integration. The architecture of the CNNs allows to connect the three steps sequentially and to fully analyze the raw signals in one block. A test performed with the algorithm shows that the CNN analysis is about 65 faster than the traditional deconvolution codes used for the same tasks. This opens a possibility to have in a near future an almost real-time analysis during an experiment. \par
The CNNs were able to learn complex signal processing with relative errors of less than 6\%. Although one of the CNNs performs a signal deconvolution to extract a multiple Gaussian components, the output of CNN corresponds to a single  $(512\times 1)$ array. This problem can affect the peak detection for some Gaussian functions that are not pronounced in the deconvoluted spectrum. In the future, we plan to extend the CNN step to provide a set of $(512\times n)$ arrays that can bring the information of the full signal components.

\section*{Acknowledgments}
This work was financially  supported by Fundaç\~ao de Amparo a Pesquisa do Estado de S\~ao Paulo (FAPESP) under Grant Nos.~2018/04965-4, 2016/17612-7,  2019/07767-1 and 2015/22308-2. G.F.F. thanks to Comissão Nacional de Energia Nuclear (CNEN) for the financial support within the MSc. scholarship program. The authors acknowledges support by the project INCT-FNA (464898/2014-5). We thank  the AT-TPC collaboration for the experimental data used for this work.

\bibliography{bibliography}  

\end{document}